# The ferroelectric polarization of $Y_2CoMnO_6$ aligns along the b-axis: the first-principles calculations

C. Y. Ma,[a] S. Dong,[b] P. X. Zhou,[a] Z. Z. Du,[a] M. F. Liu,[a] H. M. Liu,[a] Z. B. Yan[a] and J.-M. Liu[a,c]

[a.] *Laboratory of Solid State Microstructures and Innovation Center of Advanced Microstructures, Nanjing University, Nanjing 210093, China*
[b.] *Department of Physics, Southeast University, Nanjing 211189, China*
[c.] *Institute for Advanced Materials, South China Normal University, Guangzhou 510006, China*

Double-perovskite $A_2BB'O_6$ oxides with magnetic B and B′ ions and E*-type antiferromagnetic order (E*-AFM, i.e. the ↑↑↓↓ structure) are believed to exhibit promising multiferroic properties, and $Y_2CoMnO_6$ (YCMO) is one candidate in this category. However, the microscopic origins for magnetically induced ferroelectricity in YCMO remain unclear. In this work, we perform detailed symmetry analysis on the exchange striction effect and lattice distortion, plus the first-principles calculations on YCMO. The E*-AFM state as the ground state with other competing states such as ferromagnetic and A-antiferromagnetic orders is confirmed. It is revealed that the ferroelectricity is generated by the exchange striction associated with the E*-AFM order and chemically rdered Mn/Co occupation. Both the lattice symmetry consideration and first-principles calculations predict that the electric polarization aligns along the *b*-axis. The calculated polarization reaches up to 0.4682 $\mu C/cm^2$, mainly from the ionic displacement contribution. The present work presents a comprehensive understanding of the multiferroic mechanisms in YCMO and is of general significance for predicting emergent multiferroicity in other double-perovskite magnetic oxides.

## Introduction

Materials of multifold functionalities such as ferroelectricity, magnetism, and ferroelasticity have been receiving everlasting attention, driving substantial efforts in miniaturization, integration, and high-density storage technologies of devices/systems made of these materials.[1] One of the driving forces along these lines is exploration of multiferroic materials in which multifold ferroic functionalities coexist and inter-couple, promising favored application potentials. Following the scheme proposed by Khomskii in 2009, multiferroic materials can be categorized into two classes.[2] Type-I multiferroics exhibit good ferroelectricity and ferromagnetism but weak magnetoelectric (ME) coupling, and one example is $BiFeO_3$ whose room temperature (*T*) ferroelectric (FE) polarization (*P*) reaches up to 100 $\mu C/cm^2$ and the FE Curie point ($T_C$) up to 1103 K.[3] In general sense, type-II multiferroics exhibit strong ME coupling since the ferroelectricity originates from the spin-relevant interactions, but their $T_c$'s are quite low in most cases and the generated *P*'s are two or three orders of magnitude smaller than those of type-I multiferroics. The representatives are Tb(Dy)$MnO_3$,[4-7] $Ni_3V_2O_8$,[8,9] $MnWO_4$.[10,11] Within certain range of temperature, some of these materials have noncollinear spin ordered structures which may contribute nonzero *P*'s, this was proposed by the well-known KNB model and the Dzyaloshinskii-Moriya interaction (DMI) mechanis.[12,13] The core physics is the relativity correction of the spin exchange by the spin-orbit coupling. Therefore, the as-generated *P*'s are small with low $T_C$'s , for exanple a remnant polarization of ~ 0.06 $\mu C/cm^2$ and $T_C$ ~28 K for $TbMnO_3$.[5]

Very different from those noncollinear spin-ordered materials, specifically collinear spin-ordered materials represent another sub-group of the type-II multiferroic materials, where nonzero *P* can be generated via the symmetric exchange striction mechanism.[14] Because the exchange striction can be much stronger than the DMI mechanism, much larger *P* and higher $T_C$ can be expected in these sub-grouped materials. The well-known examples include orthorhombic $YMnO_3$, $HoMnO_3$, and $Ca_3Co_{2-x}Mn_xO_6$ (CCMO at *x* ~ 0.96).[14-20] Taking $YMnO_3$ for illustration, in single-crystal film the E-type antiferromagnetic (E-AFM) was observed below 35 K with a saturation polarization of 0.8 $\mu C/cm^2$.[19] This E-AFM order shows the ↑↑↓↓ spin alignment along the in-plane [110] direction, as shown in Fig. 1a, where the Mn spins are assumed to align along the Mn-O-Mn chain direction for simplification. If the coordinate of one magnetic ion (Mn) is taken as reference point, the Mn-O-Mn bond angle will be increased if the two Mn spins are parallel and it will be decreased if the two spins are antiparallel due to the symmetric exchange striction. This effect shifts the O ions along the direction normal to the Mn-O-Mn chain, resulting in a net *P*.

Similar ferroelectricity generation can be more generally illustrated in CCMO, as schematically shown in Fig. 1b where the O ions are ignored.[20] The Mn and Co ions stack alternatively along the *c*-axis, forming the Ising-like ↑↑↓↓ spin chains below *T* ~ 16.5 K. One may coin this spin structure as the E*-AFM order. Due to the symmetric exchange striction, the spin-parallel Co and Mn ions close to each other and the spin-antiparallel Co and Mn ions apart from each other, generating a new *P* ~ 0.009 $\mu C/cm^2$ along the *c*-axis at *T* ~ 2 K. This value is still small and the $T_C$ remains low either, but the strong exchange striction in such ↑↑↓↓ spin structures stimulates substantial effort for additional materials with improved ferroelectricity.

Along this line, attention has recently been directed to those double-perovskite oxides with complex lattice distortions and competing interactions. One prominent progress was reported by Rondinelli *et al*. who predicted high-$T_C$ ferroelectricity in some AA′$B_2O_6$ perovskite oxides where polarizations are generated by the coherent rotation and tilting of the stack-ordered octahedral units.[21] On the other hand, $Y_2NiMnO_6$ was first predicted to have a ferroelectric ground state with the E-AFM structure on the *ab*-plane, and then a polarization about 0.0035 $\mu C/cm^2$ was measured at low temperature.[22,23] It is noted that there are a number of oxides in this category where the B site may accommodate different magnetic ions.

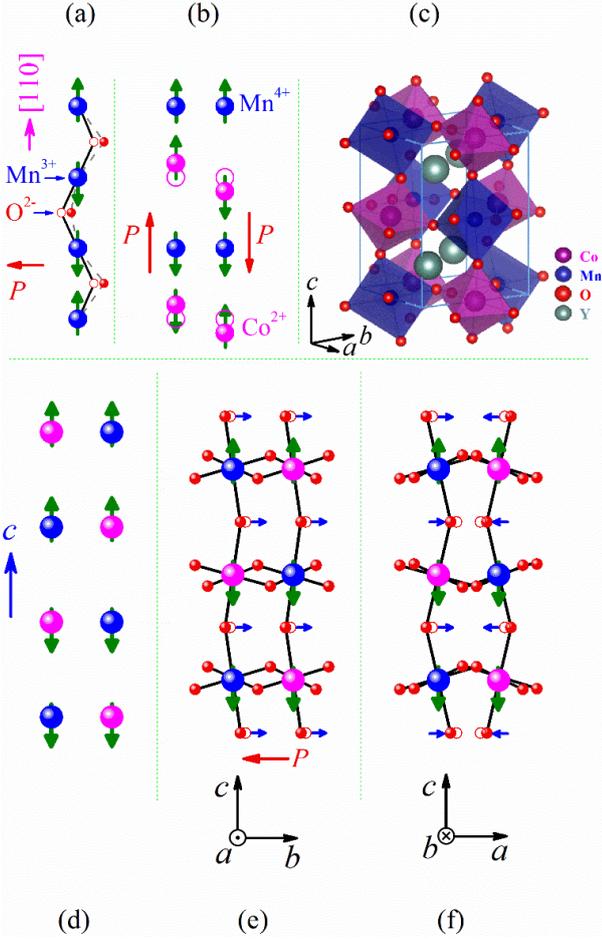

**Fig.1** (a) Symmetric exchange striction and transverse electric polarization generation of an E-AFM Mn-O-Mn chain (here orthorhombic YMnO₃ is taken as an example). (b) Symmetric exchange striction of an ↑↑↓↓ Mn-Co chain and longitudinal electric polarization in CCMO. (c) The unit cell of YCMO. (d) Spin configuration along the *c*-axis of two neighboring Mn-Co chains for YCMO. (e) & (f) The symmetric exchange striction induced O ionic displacements of two neighboring Mn-Co spin chains projected on the *bc*-plane and *ab*-plane respectively, where small red solid and open dots represent the O ionic positions before and after the displacemet .

Recent experiments revealed the ferroelectricity of other double perovskites, e.g. R₂CoMnO₆ (R = Lu, Y, Sm), Y₂MnCrO₆.[24-27] The findings in YCMO (the $T_C$ ~80 K is appreciated) at least hint promising potentials in such double-perovskite oxides A₂BB′O₆. Besides, these materials may offer other extra-interested effects, such as meta-magnet-like magnetization steps,[28,29] multicaloric effect,[30] spin-glass-like behaviors, dielectric relaxation,[31] and exchange bias.[32] In this work, we pay attention to the multiferroicity of YCMO.

## Lattice symmetry consideration

The YCMO exhibits the monoclinic structure with crystal group of $P2_1/n$ at room temperature. The lattice parameters are $a_0 = 5.2322(2)$ Å, $b_0 = 5.5901(2)$ Å, $c_0 = 7.4685(3)$ Å, $\alpha = \gamma = 90.00°$, and $\beta = 89.92(4)°$, with a schematic drawing of the lattice structure in Fig. 1c.[25] It can be viewed as a half-substituted YMnO₃ at Mn site by Co ion, distorting the Mn(Co)O₆ octahedra coherently. The Co²⁺ ions and Mn⁴⁺ ions occupy respectively the 2*c*(0, ½, 0) and 2*d*(½, 0, 0) Wyckoff positions. This substitution lowers the lattice symmetry from orthorhombic lattice into monoclinic lattice.

The E*-AFM order starts to form at $T_N$ ~ 80 K with the Co/Mn spins all aligning along the *c*-axis.[25] The magnetization *M* as a function of *T* shows a ferromagnetic (FM) like transition at *T* ~ 70-80 K,[25,28,33] but the saturated *M* at low *T* (*M* ~ 0.7 $\mu_B$/f.u. under magnetic field *H* ~100 Oe)[25] is much smaller than the full ferromagnetically aligned moment, suggesting that the magnetic state is not a full FM state but an E*-AFM state probably with spin-canted moment instead. In fact, a coexisting E*-AFM and FM state with dominant E*-AFM phase was claimed, which also explains the weak FM transition.[25,28] Due to the existence of Mn and Co ions, it is expected that the magnetic ground state may not be unique but from competing magnetic states. One of the major issues in this work is to sort out of the ground state.

The measured polarization for polycrystalline YCMO samples at low *T* is ~ 0.0065 μC/cm².[25] The similarity between YCMO and CCMO allows a prediction of polarization along the *c*-axis. Nevertheless, a detailed analysis suggests distinct difference between YCMO and CCMO in terms of ferroelectricity origin. For CCMO, it is found that the *ab*-planes are fully occupied by Co or Mn ions, and the Mn-occupied planes and Co-occupied planes stack alternatively, leading to the alternative occupations of the Co and Mn on each chain along the *c*-axis. In addition, the Mn-Co-Mn-Co chains along the *c*-axis can be approximately treated as the quasi-one-dimensional spin chains since the in-plane exchange interactions (or say the inter-chain exchanges) are weak with respect to the intra-chain exchanges. For YCMO, the *c*-axis Mn/Co chain is similar to that in CCMO. However, each in-plane is neither fully occupied by Co nor fully by Mn, instead by Co/Mn alternative occupation, as shown in Fig. 1c. The electric dipoles generated by the symmetric exchange striction between two adjacent Co-Mn layers (along the *c*-axis) would have opposite directions, leading to polarization cancellation. In particular, the in-plane exchanges are comparable with those along the *c*-axis. These facts suggest a different mechanism for ferroelectricity generation in YCMO from that in CCMO. It also raises a critical question: does polarization *P* in YCMO align along the *c*-axis? This is the second major issue in this work.

Using the Glazer notation, YCMO was reported to follow the $a^-a^-c^+$ mode for the oxygen octahedra rotation.[34] The projections of the lattice distortion onto the *bc*-plane and *ac*-plane are different. We look at two neighboring *c*-axis chains, as shown in Fig. 1d. The projection details on the *bc*- and *ac*-planes are schematically drawn in Fig. 1e and f, respectively. As shown in Fig. 1e, the small red solid dots represent the O ions which deviate from the high-symmetry positions due to the oxygen octahedra rotation. In this case, the ↑↑↓↓ spin alignment along the *c*-axis drives these O ions to further shift via the symmetric exchange striction, as shown by the small open dots. It is interested to note that all these O ions on the two neighboring chains shift coherently along the +*b*-axis, generating a net *P* along the -*b*-axis. However, as shown in Fig. 1f, the ↑↑↓↓ spin alignment along the *c*-axis also drives the O ions on the chain to shift via the exchange striction. In this case, the O ions on the two neighboring chains shift towards opposite directions, leading to a cancellation of the local electric dipoles along the ±$a$-axis.

The abovementioned model prediction suggests that the polarization $P$ in YCMO aligns along the $b$-axis instead of the $c$-axis, different from earlier predictions. Surely, one understands that the polarizations along the $a$-axis and $c$-axis may not be exactly cancelled. And more important is that a quantitative check of this model prediction should be of general significance for understanding the multiferroic behaviors of the whole $A_2BB'O_6$ family. First-principles calculation based on density functional theory (DFT) is a powerful tool for study of the structural, electronic, magnetic and ferroelectric properties of materials.[35] Even though in principle the DFT calculation can only deal with the zero-temperature ground state, it remains useful to understand the physics of materials, including those type-II multiferroics.[16-18,36-39] In this work, we pay attention to the full-scale first-principles calculations of the lattice and electronic structures of YCMO not only for checking the above prediction.

## First-principles calculations

We perform the density functional theory calculations using the VASP (Vienna *ab initio* simulation package) code on the basis of the projector augmented waves scheme (PAW).[40,41] The Perdew-Burke-Ernzerhof functional version of the generalized-gradient approximations (GGA) with on-site Coulomb interaction (within the Dudarev's approach for the Co and Mn orbitals are used for ionic relaxation and polarization calculations.[42,43] We explicitly treat eleven valence electrons for Y ($4s^2 4p^6 4d^1 5s^2$), thirteen for Mn ($3p^6 3d^5 4s^2$), nine for Co ($3d^7 4s^2$), and six for O ($2s^2 2p^4$).

To accommodate the magnetic structure, a supercell, constructed by doubling the unit along $c$-axis (Fig. 1c), is used for constructing the E*-AFM order. The Γ-centered Monkhorst-Pack $k$-mesh of 6×6×3 is used for the monoclinic calculations. The global break condition for the electronic self-consistency is $10^{-6}$ eV. In order to check the ground state, four types of magnetic orders are under consideration here: FM order, E*-AFM order, A-AFM order, and FIM* order (↑↑↑↓ spin alignment), as shown in Fig. 2a, noting that all the four orders have the in-plane FM alignment.

Because these is no low-temperature experimental structural information available, the experimental room-temperature lattice constants are adopted in our calculation. the atomic positions are fully relaxed with the four candidate magnetic configurations, respectively. The criterion of relaxation is 0.005 eV/Å. Given the magnetic ground state, the ferroelectric polarization is calculated using the standard Berry-phase method.[44] The FM state, with a nonpolar structure, is taken as the paraelectric reference.

## Results and discussion

### Magnetic ground state

Because experimentally determined on-site Coulomb potential $U$ value is not available, we test the dependence of the total energies on different $U_{eff} = U - J$ values from 1.0 eV to 7.0 eV for Co and Mn $d$-orbitals. Herein we choose the equal value of $U_{eff}$ for Co and Mn between each individual calculation mostly for simplification. Moreover, it is reasonable to treat Co and Mn $d$-orbitals equally

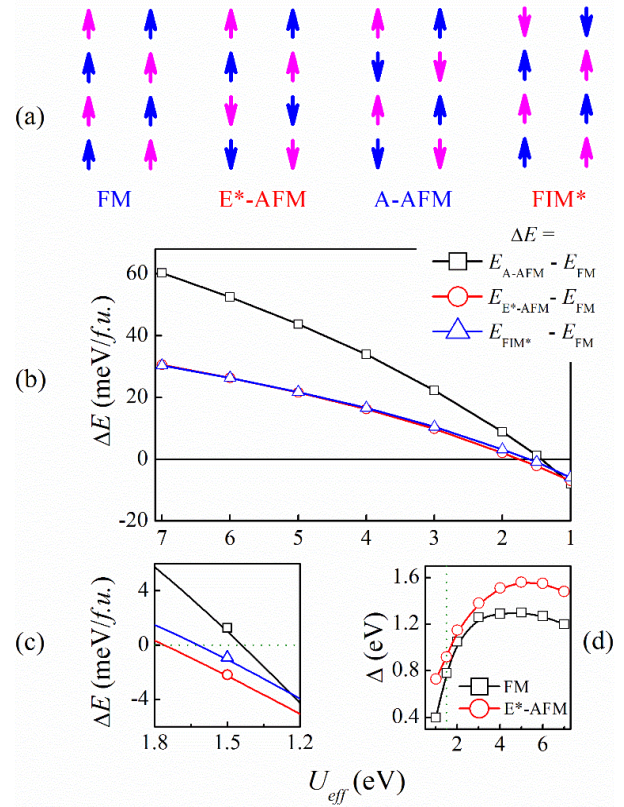

**Fig. 2** (a) Four spin orders for YCMO: FM, E*-AFM, A-AFM, and FIM*, with the pink dots for Co and blue dots for Mn. (b) Calculated total energy differences ΔE of the E*-AFM, A-AFM, and FIM* states from the FM states, as a function of $U_{eff}$. (c) The amplified local region of (b) between $U_{eff}$ = 1.2 ~ 1.8 eV. (d) Calculated band gap Δ as a function of $U_{eff}$ for the FM and E*-AFM state respectively.

because the effective $U$ values for Co and Mn are quite close in earlier reported DFT calculations. For example, for CCMO, Wu *et al.* chose $U$ = 5.0 (4.0) eV for Co (Mn) and $J$ = 0.9 eV for both Co and Mn,[39] Zhang *et al.* used the same $U_{eff}$ = 1.1 eV for Co and Mn.[37] The dependence of the band structures and magnetic ground state can be checked more comprehensively when the $U_{eff}$ varies over a broad range from 1.0 eV to 7.0 eV.

The calculated data are presented in Fig. 2b where the energy difference between the assigned spin order and the FM order is plotted as a function of $U_{eff}$. It is seen that the FM state is the ground state at $U_{eff}$ > 1.8 eV while the ground state at $U_{eff}$ < 1.8 eV favors the E*-AFM order. Earlier calculations did reveal that a varying $U_{eff}$ influences the magnetic ground state of multiferroics, such as YMnO$_3$ compound and CCMO.[37,38] Because the E*-AFM order is the experimentally determined ground state, one may focus on the range of $U_{eff}$ from 1.0 eV to 2.0 eV. This choice is reasonable. First, the four magnetic states have comparable energies in this range and their differences are less than 5.0 meV/$f.u.$, suggesting the possible phase coexistence, in particular the coexistence of the FM state and E*-AFM state, as claimed experimentally. Second, the electronic structure calculations show a clear gap Δ associated with the E*-AFM state, and the Δ($U_{eff}$) dependence is presented in Fig. 2c, suggesting a reasonable $U_{eff}$ value less than 5.0 eV. Third, so far no measured Δ value for YCMO has been available, and one may refer to the Δ data of YMnO$_3$ and YCoO$_3$, which are 1.55 eV and 1.80 eV (DFT calculated data).[45,46] Therefore, $U_{eff}$ ~ 1.5 eV for YCMO is an acceptable value considering the under-estimation of the gap by the GGA+U scheme. Indeed, the ground state is the E*-AFM state at $U_{eff}$ ~ 1.5 eV. We fix this value for other calculations hereafter.

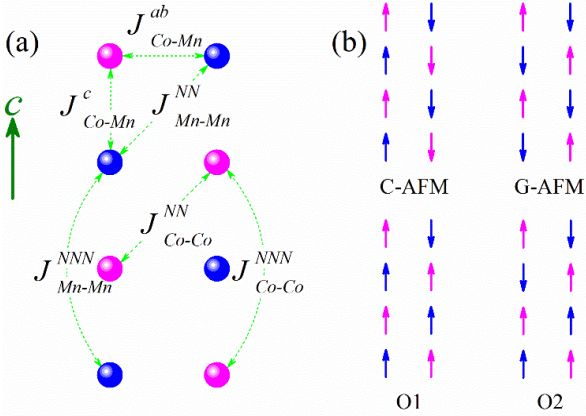

**Fig. 3** Definition of exchange parameters (a) and additional four types of spin orders (b) for YCMO. The O1 and O2 orders are artificially constructed.

To further understand why the E*-AFM order is the ground state, we deal with the Heisenberg model with the simplest Hamiltonian:

$$H = \sum_{ij} J_{ij} s_i \cdot s_j / |s_i||s_j| ,  \qquad (1)$$

where $J_{ij}$ is the exchange interaction for spin pair <ij>, given the normalized moments. It is known that such double-perovskite structure accommodates multifold interactions. For the YCMO, at most six exchange parameters can be taken into account for a full-scale model description of magnetic structure. These exchanges are indicated in Fig. 3a, where $J_{B-B'}^R$ denotes the exchange between sites B and B′ in the nearest-neighboring (R=NN), next- nearest-neighboring (R=NNN), in-plane neighboring (R=ab), or out-of-plane neighboring (R=c) configuration. to obtain the six exchange parameters from the first-principles calculations, one has to consider a number of spin configurations and calculate the corresponding total exchange energy $E_{ex}$ so that these parameters can be extracted. Besides the FM, A-AFM, and E*-AFM structures considered above, we construct additional four types of spin orders, i.e. the C-AFM, G-AFM, O1, and O2, where the latter two orders are artificially assumed only for convenience but unavailable. The spin structures of these orders are schematically shown in Fig. 3b for reference.

Table 1 The exchange energy $E_{ex}$ of the six spin structures (orders) and the extracted six exchange parameters $J_{ij}$, setting $E_{ex}$=0 for the FM order.

| Spin orders | $E_{ex}$ (meV) | Exchange $J_{ij}$ | Value (meV) |
|---|---|---|---|
| A-AFM | 4.51 | $J_{Co-Mn}^c$ | 0.678 |
| E*-AFM | 2.68 | $J_{Co-Mn}^{ab}$ | -5.113 |
| C-AFM | 178.96 | $J_{Co-Co}^{NN}$ | 0.398 |
| G-AFM | 152.77 | $J_{Mn-Mn}^{NN}$ | -0.878 |
| O1 | 85.82 | $J_{Co-Co}^{NNN}$ | -0.629 |
| O2 | 49.93 | $J_{Mn-Mn}^{NNN}$ | 0.576 |

In consequence, the exchange energy terms of the two formulas (f.u.) for the seven spin orders can be written as follows:

$$\begin{cases} E(FM) = 8J_{Co-Mn}^c + 16J_{Co-Mn}^{ab} + 16J_{Co-Co}^{NN} + 16J_{Mn-Mn}^{NN} \\ \qquad\qquad + 4J_{Co-Co}^{NNN} + 4J_{Mn-Mn}^{NNN} \\ E(A\text{-}AFM) = -8J_{Co-Mn}^c + 16J_{Co-Mn}^{ab} - 16J_{Co-Co}^{NN} - 16J_{Mn-Mn}^{NN} \\ \qquad\qquad + 4J_{Co-Co}^{NNN} + 4J_{Mn-Mn}^{NNN} \\ E(C\text{-}AFM) = 8J_{Co-Mn}^c - 16J_{Co-Mn}^{ab} - 16J_{Co-Co}^{NN} - 16J_{Mn-Mn}^{NN} \\ \qquad\qquad + 4J_{Co-Co}^{NNN} + 4J_{Mn-Mn}^{NNN} \\ E(G\text{-}AFM) = -8J_{Co-Mn}^c - 16J_{Co-Mn}^{ab} + 16J_{Co-Co}^{NN} + 16J_{Mn-Mn}^{NN} \\ \qquad\qquad + 4J_{Co-Co}^{NNN} + 4J_{Mn-Mn}^{NNN} \\ E(E^*\text{-}AFM) = 16J_{Co-Mn}^{ab} - 4J_{Co-Co}^{NNN} - 4J_{Mn-Mn}^{NNN} \\ E(O1) = 16J_{Co-Co}^{NN} + 4J_{Co-Co}^{NNN} - 4J_{Mn-Mn}^{NNN} \\ E(O2) = 4J_{Co-Mn}^c + 8J_{Co-Mn}^{ab} + 16J_{Co-Co}^{NN} + 4J_{Co-Co}^{NNN} \end{cases} \qquad (2)$$

In comparison, the energy of the FM order is taken as the zero-point reference and we calculate the energy differences of other six spin orders from the FM order, given the pre-condition that these energies are calculated by fixing all the atoms on the high-symmetry positions of the FM order. The calculated total energy terms of these spin orders other than the FM order and extracted six exchange parameters are listed in Table 1. The two nearest-neighboring Co-Mn exchanges $J_{Co-Mn}^c$ = 0.678 meV and $J_{Co-Mn}^{ab}$ = -5.113 meV, indicating the out-of-plane AFM exchange and in-plane FM exchange. The two exchanges are dominant over other four exchanges, and in particular one sees $|J_{Co-Mn}^{ab}| > |J_{Co-Mn}^c|$, suggesting the even stronger in-plane exchange than the out-of-plane exchange, which is distinctly different from the situation of CCMO. From this point of view, the claimed phase coexistence, such as the A-AFM/FM/E*-AFM phase coexistence, becomes physically reasonable. Even though, upon the full lattice relaxation, the calculations show that the E*-AFM order has the lowest exchange energy, while the C-AFM, G-AFM, O1, and O2 orders are most likely unavailable. These predictions are consistent qualitatively with experimental observations, thus providing a physical basis for the magnetically induced ferro-electricity in YCMO.

According to these exchange interactions, the transition temperature can be roughly estimated in the mean-field level. Taking the nearest-neighboring interactions into consideration, the Néel temperature for the E*-AFM phase is $k_B T_N = 4J_{Co-Mn}^{ab}/3$, where $k_B$ is the Boltzmann constant.[47] The estimated $T_N$ = 79 K is in good agreement with experimental temperature 80 K at which the spontaneous polarization first emerges.[25]

**Band structures**

complimentary to the magnetic structure calculations, the band structures and charge distributions are computed, particularly focusing on the FM order and E*-AFM order. The band structures for the two phases are plotted in Fig. 4a and b. The blue and red lines in Fig. 4b mark the spin-up and spin-down states, respectively. It is revealed that the FM and E*-AFM phases exhibit a band gap of 0.92 eV and 0.78 eV respectively, indicating the insulating-semiconducting behaviors. To uncover the orbital contributions of Mn and Co ions, we extract the spin-resolved density of states (DOS) of the 3d-orbitals for Mn and Co ions, as shown in Fig. 4c~e where the blue lines for the spin-up states and the red lines for the spin-down states. The 2$p$ states of O ions are given in Fig. 4f. It is shown that the valence band maximum (VBM) is contributed mainly by the spin-down $t_{2g}$-orbitals of Co electrons while the conduction band minimum (CBM) comes from the spin-up $e_g$-orbitals of Mn electrons and the spin-down $t_{2g}$-orbitals of Co electrons together. The strong hybridization between Mn/Co 3$d$ electrons and O 2$p$ electrons can be clearly identified in the range from -7.5 eV to -1.5 eV. Furthermore, only three spin-up $t_{2g}$ electrons for Mn are revealed, but there are five occupied spin-up orbitals ($t_{2g}$ and $e_g$) and two occupied spin-down orbitals for Co electrons. This confirms the Mn$^{4+}$ valence state and the Co$^{2+}$ valence state. By the way, the small bump around ~ -1.0eV in the Mn spin-up DOS, as marked by the pink shadow, includes the contributions from the Co spin-up $e_g$ electrons and the O 2$p$ electrons. This feature, in spite of weak, implies small amount of charge transfer from the the FM order and the total moment per unit is ~ 5.6 $\mu_B$, close to expected value of ~ 6.2 $\mu_B$. Finally, it is noted that the moment $M_{Mn}$ is the highest in the FM state and the lowest in the A-AFM state, while it is opposite for the $M_{Co}$ which is the highest in the A-AFM state and the lowest in the FM state, an indirect evidence for the charge transfer between the Mn and Co ions.

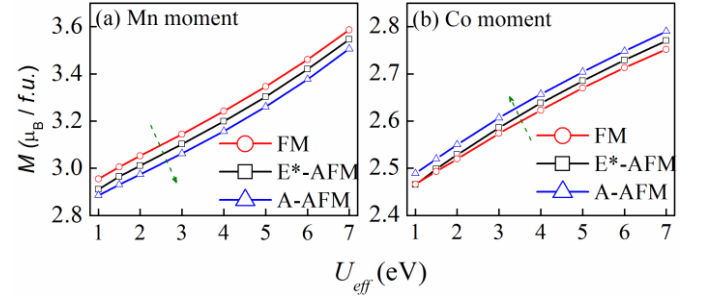

**Fig. 5** Calculated magnetic moment for Mn (a) and Co (b) in the FM, E*-AFM, and A-AFM states as a function of $U_{eff}$.

### Exchange striction effect

Then, we investigate the microscopic mechanism for polarization $P$ in YCMO. Usually, polarization contains two components, i.e. the ionic displacement induced polarization $P_{ion}$ and electronic contribution $P_{ele}$. They may enhance or cancel with each other, depending on the electronic structure and lattice distortion. We evaluate the data of all ionic positions on the Mn-O-Co chain along the $c$-axis, given different magnetic phases, so that the ionic displacements can be analyzed. The parameters assigned to define the chain geometry are presented in Fig. 6c. The inter-ionic distance and Mn-O-Co bond angle are denoted by $d$ and $\varphi$, where superscript '$p$' or '$ap$' stands for parallel spin pair or antiparallel spin pair, and subscript '$i$-$j$' labels the two ions $i$ and $j$. For example, $d^{ap}_{Co-Mn}$ stands for the separation of a neighboring Co-Mn pair whose spins are antiparallel. We count these distances and angles for the E*-AFM, FM, and A-AFM chains and compare them at different $U_{eff}$ values. The results are summarized in Fig. 6a and b, respectively, where the vertical dash line marks the position of $U_{eff}$ = 1.5 eV.

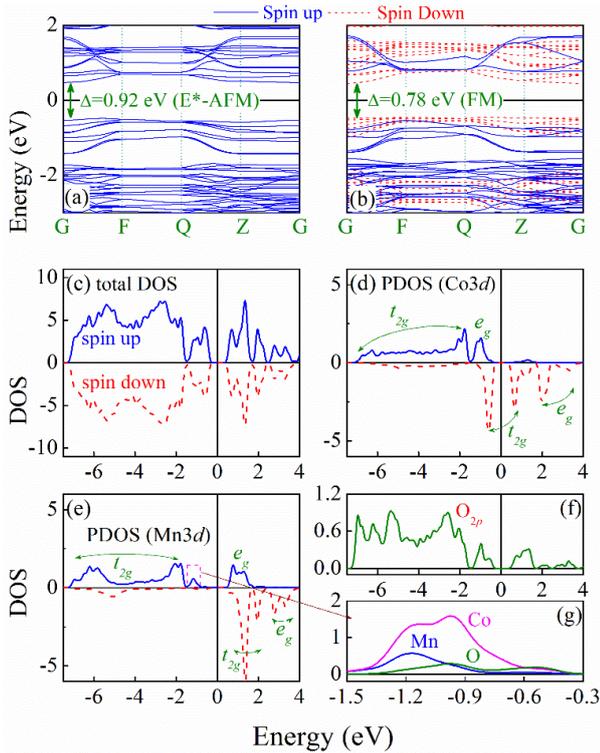

**Fig. 4** Calculated band structures for the E*-AFM state (a) and FM state (b) with the direct band gap Δ. The spin-resolved total DOS spectrum is plotted in (c). The partial DOS for the Co 3$d$ and Mn 3$d$ orbitals are plotted in (d) and (e). The O 2$p$ orbital DOS is presented in (f). The partial DOS spectra for Mn 3$d$, Co 3$d$, and O 2$p$ from -1.5 eV to -0.3 eV are shown in (g).

Co 3$d$-orbitals to the Mn 3$d$-orbitals via the inter-bridged O 2$p$-orbitals, as more clearly shown in Fig. 4g.

The above mentioned charge transfer can be partially confirmed by the calculated Mn and Co moments ($M_{Mn}$, $M_{Co}$), as shown in Fig. 5a and b respectively, where the $M_{Mn}$ and $M_{Co}$ for the FM, E*-AFM, and A-AFM orders as a function of $U_{eff}$ are plotted. For the simplest case, the Mn$^{4+}$ and Co$^{2+}$ would have the same moment. However, the calculated $M_{Mn}$ is ~ 0.4 $\mu_B$ greater than $M_{Co}$, which is partially due to the charge transfer from Co$^{2+}$ to Mn$^{4+}$, and other contributions may exist as well. Both $M_{Mn}$ and $M_{Co}$ show a linear increase with $U_{eff}$ which is reasonable considering the electron correlation induced localization effect and enhanced energy penalty for unoccupied states. The moments at $U_{eff}$ = 1.5 eV, the value taken for the present calculations, are $M_{Mn}$ ~ 3.05 $\mu_B$ and $M_{Co}$ ~ 2.5 $\mu_B$ for

As shown in Fig. 6a, for the FM chain and A-AFM chain, the Mn-Co separations are identical, i.e. $d^{ap}_{Co-Mn}$(FM)= $d^{ap}_{Co-Mn}$(A-AFM), which is easy to understand. Taking this separation as a reference, the E*-AFM chain does have the elongated (shortened) distance for the spin-parallel (antiparallel) Co-Mn pair, demonstrating the symmetric exchange striction effect. This exchange striction is weakened with increasing $U_{eff}$, due to the more localized charge distribution at the larger $U_{eff}$. At $U_{eff}$ = 1.5 eV, $d^p_{Co-Mn}$(E*-AFM) and $d^{ap}_{Co-Mn}$(E*-AFM) are ~ 0.015 Å longer and shorter respectively than $d^p_{Co-Mn}$(FM) or $d^{ap}_{Co-Mn}$(A-AFM), significant exchange strictions, respectively. In parallel to the variations of the Co-Mn pair distances, one also observes $\varphi^p$(E*-AFM) > $\varphi^p$(FM) > $\varphi^{ap}$(A-AFM) > $\varphi^{ap}$(E*-AFM), as shown in Fig. 6b. This implies that the O ion between the spin-antiparallel Co-Mn pair will have bigger transverse shift than that between the spin-parallel Co-Mn pair, contributing the transverse electric polarization, as indicated in Fig. 6c. It is noted that these angles decrease gradually with increasing $U_{eff}$, a consequence of the more localized charge distribution at larger $U_{eff}$.

Finnally, it should be noted that the space group for relaxed structure under E*-AFM is lowered from original $P2_1/n$ to $P2_1$,

whose point group is polar and thus allows a spontaneous polarization.

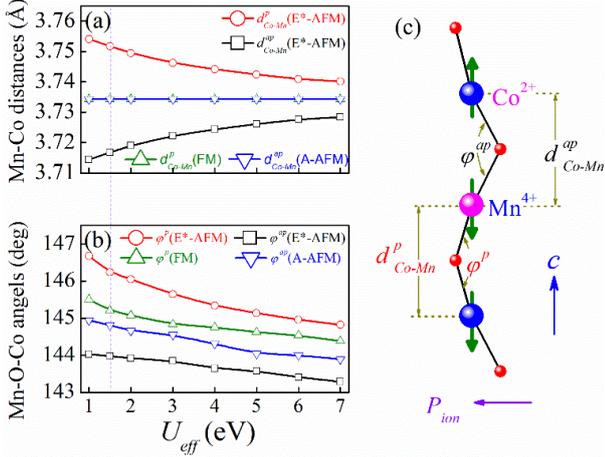

**Fig. 6** (a) Calculated Mn-Co distances ($d$) and Mn-O-Co bond angles ($\varphi$) in a Mn-Co chain along the c-axis as a function of Ueff. (b) Schematic definitions of these distances and angles. The superscripts p and ap stand for parallel and antiparallel Mn-Co spin pairs.

Table 2 Calculated total polarization $P$ and $\Delta\varphi = \varphi^p - \varphi^{ap}$ for the E*-AFM chain

| $U_{eff}$ (eV) | $\Delta\varphi$ (Deg.) | $P$ ($\mu C/cm^2$) |
|---|---|---|
| 1.0 | 2.651 | -0.4980 |
| 1.5 | 2.408 | -0.4682 |
| 2.0 | 2.265 | -0.4529 |
| 3.0 | 2.041 | -0.4086 |
| 4.0 | 1.878 | -0.3568 |
| 5.0 | 1.771 | -0.3158 |
| 6.0 | 1.692 | -0.2887 |
| 7.0 | 1.538 | -0.2582 |

**Ferroelectric polarization**

The Berry phase method is used to calculate $P$, and the calculated data at various $U_{eff}$ are listed in Table II. It is confirmed that the polarization aligns along the $-b$-axis, consistent with the above discussions on the lattice symmetry and symmetric exchange striction. The calculated $P$ is ~ -0.4682 $\mu C/cm^2$ at $U_{eff}$ = 1.5 eV and decreases roughly linearly with increasing $U_{eff}$. We show $\Delta\varphi(U_{eff})$ = $\varphi^p$(E*-AFM)-$\varphi^{ap}$(E*-AFM), the Mn-O-Co bond angle difference, which also decreases with increasing $U_{eff}$ either. The pure electronic polarization without ionic contribution is then calculated. We first impose all ions on the lattice structure of the FM phase which is paraelectric, and then the FM order is replaced by the E*-AFM order. In this case, the calculated $P_{ele}$ is ~ -0.0134 $\mu C/cm^2$ at $U_{eff}$ = 1.5 eV. To verify the results, we relax the ionic positions again with non-collinear magnetic structures and take spin-orbit coupling (SOC) into consideration. Then polarization is calculated as before. The total polarization is -0.4729 $\mu C/cm^2$ and the pure electronic contribution is -0.0082 $\mu C/cm^2$ respectively. It can be concluded that the polarization in YCMO is mainly from the ionic displacement contribution, and both $P_{ele}$ and $P_{ion}$ align along the same direction.

To verify the switch of FE polarization, the $-P$ state is also calculated, which can be simply obtained using the ↑↓↓↑ magnetic configuration. In the relaxed lattice, the ions move oppositely comparing with the ↑↑↓↓ state. The calculated polarization is 0.4686 $\mu C/cm^2$ which is almost identical to the absolute value of the ↑↑↓↓ state. In fact, similar magnetism-driven FE switching were report for many type-II multiferroics such as $YMnO_3$, $HoMnO_3$ and $BaFe_2Se_3$.[16,17,36] Furthermore, it should be noted that this polarization can be tuned by magnetic field. For example, if a strong enough field can fully suppress the ↑↑↓↓ state, the polarization can be eliminated, since the FM state is nonpolar in our calculation. Experimentally, an external field fo 5T could supress the magnitude of polarization by 10%.[25]

Another point is that the calculated $P$ is ~ 70 times larger than measured value for polycrystalline YCMO. Similar larger calculated values were also reported in a few materials with exchange striction induced ferroelectricity, such as $HoMnO_3$, $YMnO_3$, and CCMO.[16-18,38] Besides the polycrystalline state of the samples, other possible reasons for this difference include the phase separated state and possible antisite occupations of Mn and Co ions in the samples.[28,29] The effects of antisite disorder on the magnetic properties of double perovskites $Y_2CoMnO_6$ have been experimentally investigated. Antisite ions would destroy the perfect $Co^{2+}$-O-$Mn^{4+}$ interaction and create $Mn^{4+}$-O-$Mn^{4+}$ or $Co^{2+}$-O-$Co^{2+}$ interactions, even $Co^{3+}$ and $Mn^{3+}$.[28] This disorder may also lead to weak ferromagnetism, lower the critical temperatures, induce magnetization steps in the hysteresis, and corrodes the ↑↑↓↓ spin pattern and the induced polarization.[26,28] Experimentally, the percentage of antisite disorder ions can be controlled during the preparation.[26,28] Although Co and Mn lie close in periodic table, the ionic sizes of $Co^{2+}$ and $Mn^{4+}$ are quite different. For example, in the six-coordinate octahedron, $Mn^{4+}$'s size is 67 pm while that for the high spin state of $Co^{2+}$ is 88.5 pm.[48] This difference provides the possiblity to control the antisite occupancy. Studying the effects of antisite disorder is an interesting physical issue but is not the purpose of this study. Measurements on high quality YCMO single crystals are thus appreciated for checking the present predictions.

Then we look at the ionic displacements in the lattice which are presented in Fig. 7a~c, where the ($a$, $b$, $c$) coordinates of all the 20 ions in one super-unit cell in the E*-AFM phaseare plotted with respect to their coordinates in the high symmetry FM phase. These ions are categorized into the Y ions (four), in-plane O ions (eight), O ions on the Mn-O-Co chains along the $c$-axis (four), Mn ions (two), and Co ions (two). It is seen that any ion in one category can find its pair with opposite coordinate along both the $a$-axis and the $c$-axis, implying no net average ionic displacement along the $a$-axis and the $c$-axis for all the ions in one category. For instance, considering the $a$-axis coordinates of the four Y ions, one sees two of them (ions 1 and 4) have the opposite coordinate values and the other two (ions 2 and 3) have the opposite ones too, leading to the zero average coordinate of the four Y ions along the $a$-axis. However, the situation along the $b$-axis is very different. While the four Y ions and the eight in-plane O ions have almost zero average $b$-axis coordinates (in fact, both cases have very small values), all the four O ions along the $c$-axis shift along the $b$-axis and all the Mn and Co ions shift along the $-b$-axis, resulting in remarkable ionic polarization along the $-b$-axis. We employ the effective point-charge model to calculate the polarization which is ~ -0.5 $\mu C/cm^2$, consistent well with the Berry phase prediction. This also confirms that the ionic polarization in YCMO is dominant over the electronic one.

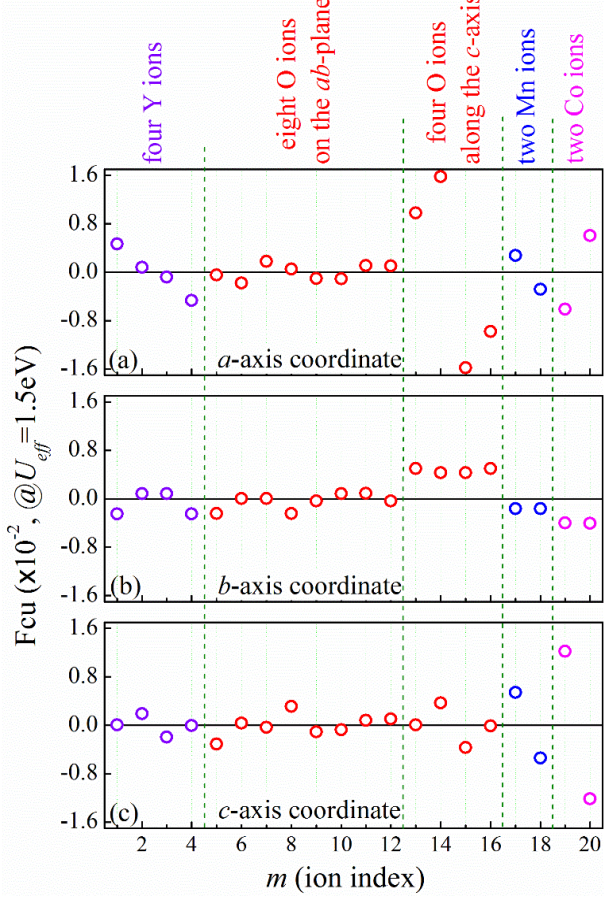

**Fig. 7** Calculated (*a*, *b*, *c*) coordinates of all the 20 ions in the super-unit cell for the E*-AFM state, with respect to the corresponding high-symmetry points of the FM state.

**Symmetry argument**

The dominant ionic polarization in YCMO has been demonstrated by our first-principles calculations plus lattice symmetry discussion, which allows an additional investigation of the origin for polarization along the *b*-axis in terms of symmetry operation language. It is known that YCMO lattice in the high symmetry FM phase has the $P2_1/n$ space group which allows two types of symmetry operations. One is the two-fold screw rotation along the *b*-axis (type-I, $P2_1$), with the screw axis labeled as the black solid dot in Fig. 8a, noting that the two Mn-O-Co chains along the *c*-axis have a $b_0/2$ out-of-plane shift from each other. The same $a_0/2$ out-of-shift shift applies to the two chains shown in Fig. 8b. The other type of operation is the glide plane perpendicular to the *b*-axis (type-II, /*n*). The following relations must be satisfied for the type-I operation of those coordination-equivalent atoms:

$$x_1 + x_2 = (n + \frac{1}{2})a_0$$
$$y_1 - y_2 = (n + \frac{1}{2})b_0 \quad , \qquad (3)$$
$$z_1 + z_2 = (n + \frac{1}{2})c_0$$

where $(x_i, y_i, z_i)$ are the coordinates of equivalent atom *i* along the (*a*, *b*, *c*)-axis and *n* is an arbitrary integer. Similarly, the type-II operation requires following equations:

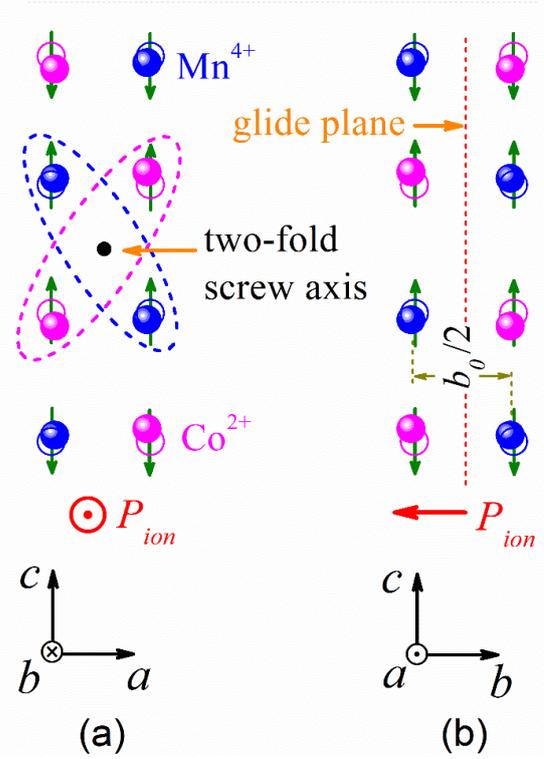

**Fig. 8** Lattice symmetry for YCMO. The two-fold screw operation is shown in (a) and the glide plane operation is given in (b) for a guide of eyes. $P_{ion}$ is the ionic polarization.

$$x_1 - x_2 = (n + \frac{1}{2})a_0$$
$$y_1 = y_2 = (n + \frac{1}{2})b_0 \quad , \qquad (4)$$
$$z_1 - z_2 = (n + \frac{1}{2})c_0$$

Eqn (3) implies the inversion symmetry of an atom along the *a*- and *c*-axis, and the transitional symmetry along the *b*-axis, while Eqn (4) implies the transitional symmetry of an atom along the *a*- and *c*-axis, and the inversion symmetry along the *b*-axis. A combination of the two types of operations guarantees the central inversion symmetry of the paraelectric/FM phase. A transition from the FM phase to the E*-AFM phase breaks the transitional symmetry along the *c*-axis, due to the exchange striction effect, which also damages the transitional symmetry along the *a*-axis and the inversion symmetry along the *b*-axis. Interestingly, it is noted that the type-I symmetry operation remains unaffected, which lower the space group to $P2_1$, as shown in Fig. 8. In consequence, all ionic displacements along the *a*-axis and *c*-axis are cancelled with each other respectively, but the displacements along the *b*-axis aren't, resulting in the net polarization along the -*b*-axis, as shown in Fig. 1e and f, 6c, and 8.

**Remarks and potential generality**

To this end, we have presented a detailed analysis of the ferroelectricity origin in YCMO with the E*-AFM ground state and $P2_1/n$ symmetry from various approaches. The central point is that the electric polarization aligns along the $b$-axis rather than the $c$-axis as well known for CCMO and other similar double-perovskite ferroelectric oxides, such as $Y_2NiMnO_6$, $R_2CoMnO_6$ (R=Lu, Y, Sm), and $Y_2MnCrO_6$. For the $A_2BB'O_6$ family, the A-site and B/B' site can be substituted by a set of different ions, such as the rare-earth La-group, Sr, Ca, Mg, Zn, Pb, and Bi etc for the A-site, and Mn, Co, Cr, Ni, Fe, and W *et al.* for the B/B' site. The O ion can be replaced by F and S etc. Rich multiferroic properties and other related magnetic properties are expected in these materials, which are associated with the lattice distortion and specific spin orders such as the E*-AFM order here. From the consideration of lattice and magnetic structures, the A-site substitution with bigger ions would be favored because higher FM transition point was reported in such cases.[49] The heterostructures as fabricated from these double-perovskite oxides as components were reported to have high ferroelectric Curie temperature.[50]

For YCMO itself, there are still several issues to be addressed. First, careful identification of the ionic displacements shown in Fig. 6 still finds very weak and even negligible polarization along the $c$-axis. A rough estimation of the $P_{ion}$ along the $c$-axis is only ~1% of that along the $b$-axis, which can be within the computational uncertainties. The underlying remains unclear to the authors. Second, the $U_{eff}$ = 1.5 eV has been taken for evaluating the ferroelectricity and it seems more direct evidence with this choice is needed. And more, the calculated polarization is still inconsistent with measured results for polycrystalline samples, while additional check with single crystal or epitaxial thin films is appealed. Third, the effects of lattice strain and external magnetic/electric field on the ferroelectricity and magnetization should deserve additional investigations.

## Conclusion

In conclusion, we have investigated in details the magnetic ground state and exchange striction effect of double-perovskite magnetic oxide YCMO ($Y_2CoMnO_6$) by performing the lattice symmetry analysis and first-principles calculations. It is found that the E*-AFM state with ↑↑↓↓ Co/Mn ordered structure is the ground state with other competing spin structures such as the FM state, A-AFM state, and O-state. The calculated but underestimated band gaps for the FM state and E*-AFM state are ~ 0.78 eV and ~ 0.92 eV, respectively. It is revealed that the symmetric exchange striction in the E*-AFM state induces an electric polarization as large as ~ 0.4682 μC/cm$^2$, aligning along the $b$-axis rather than the $c$-axis identified in $Ca_3CoMnO_6$. The ionic polarization is dominant and the electronic contribution is quite small. The possible symmetry operations for the $P2_1/n$ lattice group are discussed, from which only the $b$-axis oriented polarization is allowed. The present work sheds light on the microscopic mechanism for ferroelectricity generation in YCMO, and this mechanism may be applied to other similar double-perovskite magnetic oxides in terms of multiferroicity.

## Acknowledgement

This work was supported by the National Key Projects for Basic Researches of China (2015CB654602), the National Natural Science Foundation of China (11234005, 51332006, 51322206, 11374147), and the Priority Academic Program Development of Jiangsu Higher Education Institutions, China. The support from the High Performance Computing Center of Nanjing University is also acknowledged.